# Multiscale Entropy Analysis: A New Method to Detect Determinism in a Time Series

A. Sarkar and P. Barat

*Variable Energy Cyclotron Centre*

*1/AF Bidhan Nagar, Kolkata 700064, India*

**Abstract**

*In this letter we show that the Multiscale Entropy (MSE) analysis can detect the determinism in a time series.*

The output variables (time series) from physical systems often exhibit complex fluctuations containing information about the underlying dynamics. An important problem in the study of a time series is determining whether the time series arises from a stochastic process or has a deterministic component that is generated from chaotic dynamics having finite number of degrees of freedom. Whether a time series has a deterministic component or not in turn dictates what approaches are appropriate for investigating the time series and its generating process. In this sense detecting the determinism in a time series is very important.

Several methods of nonlinear dynamical analysis have previously been developed to detect determinism in time series [1-3]. These methods are all based on the assumption that a trajectory in the state space reconstructed from a deterministic time series behaves similarly to nearby trajectories as time evolves. Hence, a large number of data points are required to have sufficient information of the nearby trajectories to compare their future behaviors. In addition, the application of these methods can lead to spurious results for nonstationary time series.

Recently Costa et al. [4] introduced a new method, Multiscale Entropy (MSE) analysis for measuring the complexity of finite length time series. In this paper we show that the MSE method can be used to detect the determinism in a time series. The MSE method measures complexity taking into account the multiple time scales. This computational tool can be quite effectively used to quantify the complexity of a natural time series. The MSE method uses Sample Entropy (SampEn) [5] to quantify the regularity of finite length time series. SampEn is largely independent of the time series length when the total number of data points is larger than approximately 750 [5].

Recently MSE has been successfully applied to quantify the complexity of many Physiologic and Biological signals [6, 7].

The MSE method is based on the evaluation of SampEn on the multiple scales. The prescription of the MSE analysis is: given a one-dimensional discrete time series, $\{x_1,.....,x_i,....,x_N\}$, construct the consecutive coarse-grained time series, $\{y^{(\tau)}\}$, determined by the scale factor, $\tau$, according to the equation:

$$y_j^\tau = 1/\tau \sum_{i=(j-1)\tau+1}^{j\tau} x_i \qquad (1)$$

where $\tau$ represents the scale factor and $1 \leq j \leq N/\tau$. The length of each coarse-grained time series is $N/\tau$. For scale one, the coarse-grained time series is simply the original time series. Next we calculate the SampEn for each scale using the following method. Let $\{X_i\} = \{x_1,.....,x_i,.....,x_N\}$ be a time series of length N. $u_m(i) = \{x_i, x_{i+1},.....,x_{i+m-1}\}, 1 \leq i \leq N-m$ be vectors of length $m$. Let $n_{im}(r)$ represent the number of vectors $u_m(j)$ within distance $r$ of $u_m(i)$, where $j$ ranges from 1 to (N-m) and $j \neq i$ to exclude the self matches. $C_i^m(r) = n_{im}(r)/(N-m-1)$ is the probability that any $u_m(j)$ is within $r$ of $u_m(i)$. We then define

$$U^m(r) = 1/(N-m) \sum_{i=1}^{N-m} \ln C_i^m(r) \qquad (2)$$

The parameter Sample Entropy (SampEn) [5] is defined as

$$SampEn(m,r) = \lim_{N \to \infty} \left\{ -\ln \frac{U^{m+1}(r)}{U^m(r)} \right\} \qquad (3)$$

For finite length N the SampEn is estimated by the statistics

$$SampEn(m,r,N) = -\ln\frac{U^{m+1}(r)}{U^m(r)} \tag{4}$$

Advantage of SampEn is that it is less dependent on time series length and is relatively consistent over broad range of possible r, m and N values. We have calculated SampEn for all the studied data sets with the parameters m=2 and r= 0.15×SD (SD is the standard deviation of the original time series).

Costa et al. had tested the MSE method on simulated white and 1/f noises [4]. They have shown that for the scale one, the value of entropy is higher for the white noise time series in comparison to the 1/f noise. This may apparently lead to the conclusion that the inherent complexity is more in the white noise in comparison to the 1/f noise. However, the application of the MSE method shows that the value of the entropy for the 1/f noise remains almost invariant for all the scales while the value of entropy for the white noise time series monotonically decreases and for scales greater than 5, it becomes smaller than the corresponding values for the 1/f noise. This result explains the fact that the 1/f noise contains complex structures across multiple scales in contrast to the white noise. With a view to understand the complexity of deterministic chaotic data we have applied the MSE method to the following synthetic chaotic data sets.

1. Logistic Map:

$$x_{n+1} = ax_n(1-x_n) \qquad a=3.9$$

2. Henon Map:

$$x_{n+1} = 1 - \alpha x_n^2 + y_n$$
$$y_{n+1} = \beta x_n \qquad \alpha=1.4, \beta=0.3$$

3. Ikeda Map:

$$x_{n+1} = 1 + cx_n \exp(ia - \frac{ib}{1+|x_n|}) \qquad a=0.4, b=6.0, c=0.9$$

4. Quadratic Map:

$$x_{n+1} = p - x_n^2 \qquad p=1.7904$$

5. Rossler Equation:

$$\frac{dx}{dt} = -y - z$$
$$\frac{dy}{dt} = x + ay \qquad a=0.2, b=0.2, c=5.7$$
$$\frac{dz}{dt} = b + z(x - c)$$

6. Lorentz Equation:

$$\frac{dx}{dt} = -ax + ay$$
$$\frac{dy}{dt} = bx - y - xz \qquad a=10, b=28, c=8/3$$
$$\frac{dz}{dt} = -cz + xy$$

The result of the MSE analysis on the chaotic data sets together with the white noise, fractional Brownian noise (with Hurst exponent 0.7) [8] and the 1/f noise is shown in Fig. 1. It is seen that the entropy measure for the deterministic chaotic time series increases on small scales and then gradually decreases indicating the reduction of complexity on the larger scales. This trend of the variation of the SampEn with scale is entirely different from the white noise, fractional Brownian noise and the 1/f noise [4]. Moreover, the variation of the SampEn for all chaotic data sets showed a similar behavior. This establishes the fact that the MSE analysis can be used to detect the determinism in a time series.

In conclusion we have showed that the Sample Entropy is an important statistic for detecting determinism in a time series.

**Figure caption**

Fig. 1 MSE analysis of the various simulated chaotic data and white noise, fractional Brownian noise, 1/f noise each with 20000 data points.

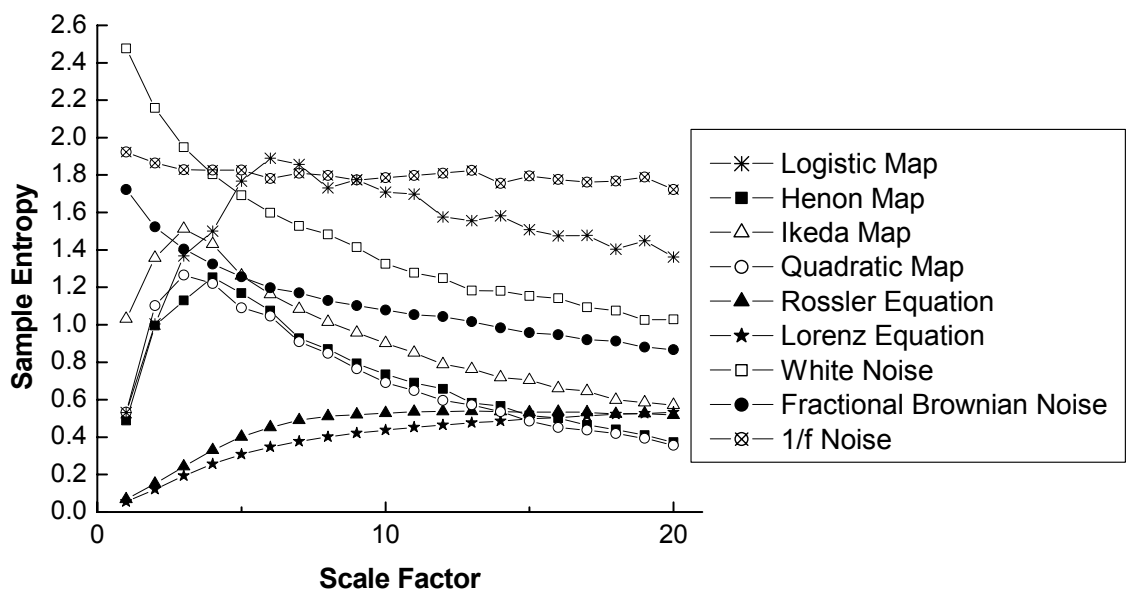

Fig. 1